\documentclass{ws-procs9x6-cpt16}
\begin{document}

\newcommand{\refeq}[1]{(\ref{#1})}
\def\etal {{\it et al.}}

\title{Ultra-High Energy Astrophysical Neutrino Detection, and the Search for Lorentz-Invariance Violations}

\author{J.C.\ Hanson}

\address{Center for Cosmology and AstroParticle Physics\\
The Ohio State University, Columbus, OH 43219, USA}

\begin{abstract}
A growing class of ultra-high energy neutrino observatories based on the Askaryan effect and Antarctic ice is able to search for Lorentz-invariance violation.  The ARA, ARIANNA, ANITA, and EVA collaborations have the power to constrain the Standard-Model Extension by measuring the flux and energy distribution of neutrinos created through the GZK process.  The future expansion of ARA, at the South Pole, pushes the discovery potential further.
\end{abstract}

\bodymatter

\section{The GZK Process and EeV neutrinos at the Earth}

Ultra-high energy neutrino (UHE-$\nu$) observations are a long-desired achievement in astroparticle physics.  Clues about cosmic ray origins and potential electroweak interaction measurements from $10^{16}$-$10^{19}$ eV are contained within this flux.\cite{kotera} PeV-scale neutrino observations in IceCube\cite{ice,ice2} have made possible learning about UHE-$\nu$ physics from beyond the solar system.  A UHE-$\nu$ could be produced via the GZK process, given the UHE-$p^{+}$ spectral cutoff at $10^{19.5}$ eV.\cite{TA}  The next generation of UHE-$\nu$ detectors is designed around the Askaryan effect, which produces radiated radiofrequency power.\cite{zhs,rb,arz,jch}  Antarctic ice provides a convenient medium for Askaryan radiation.\cite{icejch}  The RICE collaboration\cite{rice} began the field, and efforts such as ANITA, ARA, ARIANNA, and proposed EVA\cite{anita,ara,arianna,eva} have made progress in developing sensitivity to UHE-$\nu$ fluxes.

There is a connection between Lorentz-invariance violation (LIV) and UHE-$\nu$, through flux limits, via the Standard-Model Extension (SME).\cite{sme}  The SME includes LIV terms of varying dimension, proportional to small coefficients. LIV in the neutrino sector could modify the UHE-$\nu$ spectrum at Earth by introducing vacuum energy loss.\cite{gorham}  The UHE-$\nu$ detectors can place constraints on SME coefficients.

\section{Experimental detection efforts}
\label{sec:detectors}

The Antarctic Impulse Transient Antenna (ANITA) is a balloon-borne detector with radiofrequency (RF) antennas as payload.\cite{anita}  ANITA-1 flew in 2007-2008 with 32 separate RF channels, modified in subsquent seasons (40 channels for ANITA-2 and 48 for ANITA-3).  ANITA detects man-made noise, thermal noise, and RF pulses from UHE-$p^{+}$.\cite{anitacr}  ANITA has a UHE-$\nu$ threshold $E_{th} \simeq10$ EeV, limited by RF propagation to the payload ($\simeq 35$ km).  The balloon altitude allows the detector to observe instantaneously V$_{\rm eff} \Omega \approx 100$ km$^3$ str of ice at 10 EeV (the effective volume times the viewable solid angle) \cite{anita}.

The Askaryan Radio Array (ARA) is an \textit{in situ} array of RF detectors at the South Pole.\cite{ara}  Three detectors are deployed, using AC power from the Amundsen-Scott base.  \textit{In situ} detectors lower $E_{th}$ by being $\simeq 1$ km from typical events.  For example, ARA is projected to have V$_{\rm eff} \Omega \approx 1000$ km$^3$ str at 10 EeV, and $100$ km$^3$ str at 0.1 EeV \cite{ara,otherara}.  With an analysis using 2 of 37 planned stations, ARA is already competitive with ANITA below $E_{\nu} \simeq 10$ EeV and with the IceCube high-energy analysis above $E_{\nu} \simeq 100$ EeV.  ARA is projected to detect $\simeq 100$ GZK neutrinos in 3 years.\cite{kotera,ara}

The Antarctic Ross Ice Shelf Antenna Neutrino Array (ARIANNA) is another \textit{in situ} detector, located on the Ross Ice Shelf.\cite{arianna}  The Hexagonal Radio Array (HRA) is the seven-station prototype.  The ocean/ice boundary provides a mirror for RF signal collection, boosting effective volume through increased visible solid angle.\cite{icejch} Extensive air showers have been observed \cite{corey}, and final UHE-$\nu$ sensitivity is projected to be equal to ARA.  The final array design requires a $31 \times 31$ station array, with stations separated by 1 km, from in-ice attenuation length measurements.\cite{thesis}

The ExaVolt Antenna (EVA) is a proposed balloon-borne detector with a boosted RF effective area.\cite{eva}  The balloon itself is the antenna, and technological improvements in balloon design are expected to boost flight durations.  The EVA project is currently in the proposal stage.

\section{LIV tests in the Askaryan-based neutrino experiments}
\label{sec:liv}

The SME allows for UHE-$\nu$ energy loss while propagating in a vacuum.\cite{gorham}  One example is the vacuum Chernokov effect: $\nu \rightarrow \nu e^{+} e^{-}$.  UHE-$\nu$ with relatively higher energies disappear, and an abundance of lower-energy UHE-$\nu$ appears.  The energy loss may be treated like a decay with half life $\tau_\nu$ given by 

\begin{equation}
\frac{\tau_\nu}{{\rm s}} = \tau_{CG} \left(\frac{E_{\nu}}{{\rm GeV}}\right)^{-5}\frac 1{\alpha_\nu^{3}}\ ,
\label{eq:time}
\end{equation}
with $\tau_{CG} = 6.5 \times 10^{-11} $s, and $\alpha_\nu$ being the constrained SME parameter.

\begin{figure}
\begin{center}
\includegraphics[width=0.7\hsize,angle=270]{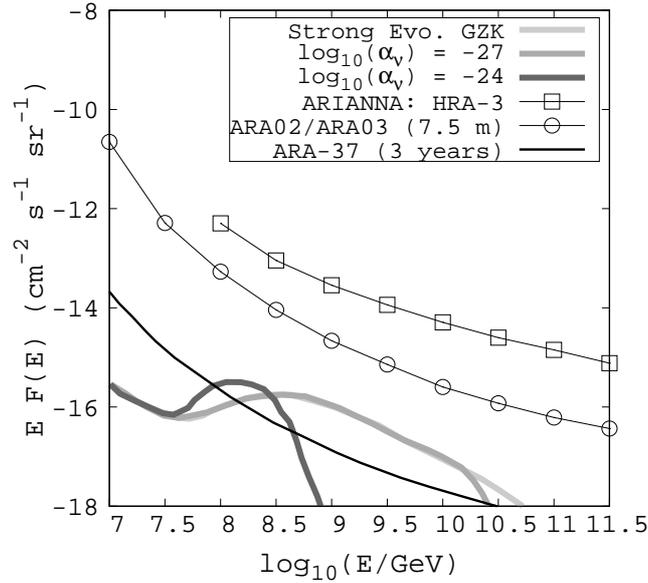}
\caption{\label{fig:money} The GZK-neutrino flux $F(E)$ times the energy $E_\nu$, vs. $E_\nu$, are shown as the thick gray lines, with $\alpha_\nu = 0$, $\log_{10}(\alpha_\nu) = -27$, and $\log_{10}(\alpha_\nu) = -24$.  
The current ARIANNA (HRA-3), ARA02/ARA03, and projected ARA-37 upper limits are shown as squares, circles, and the thin black line, respectively.  The UHE-$\nu$ spectra are adapted from Gorham $\etal$\cite{gorham}}
\end{center}
\end{figure}

Attributing the nonobservation of a GZK flux to LIV, one can place \textit{lower limits} on $\alpha_\nu$.\cite{gorham}  Figure \ref{fig:money} demonstrates the LIV modification\cite{gorham,ara} to a UHE-$\nu$ flux by a non-zero value of $\alpha_\nu$.  Observation of one UHE-$\nu$ places \textit{upper limits} on $\alpha_\nu$ lower than those from atmospheric neutrinos, due to the energies and cosmological distances of GZK models.  SME constraints will therefore be improved by enhanced ARA volume and the low-energy enhancement near $10^{17}$ eV.

\section{Future work}
\label{sec:future}

This \textit{flavor blind} UHE-$\nu$ LIV scenario could be pushed further by at least two ideas.  First, the charged leptons should cascade on the CMB/IR backgrounds, producing diffuse $\gamma$-rays.\cite{ahlers}  Combining Fermi-LAT diffuse $\gamma$-ray observations and nonobservation of UHE-$\nu$ would lead to a restricted range for $\alpha_\nu$.  Secondly, non-renormalizable, higher-dimensional SME operators also generate UHE-$\nu$ energy-loss, and the effect should increase dramatically with increasing energy, making UHE-$\nu$ an ideal messenger.\cite{datatables}  Recomputing the effective $\alpha_\nu$ from these operators would produce the first limits on those SME coefficients.  The effect should increase with energy, making UHE-$\nu$ analysis ideal.

\end{document}